\documentclass[aps,prl,superscriptaddress,letterpaper]{revtex4}
\usepackage{graphicx}
\usepackage{amssymb,amsmath}
\usepackage{float}

\usepackage{setspace}
\usepackage{parskip}
\usepackage{xcolor}
\definecolor{midnightblue}{cmyk}{1,1,0,0.1}
\definecolor{forestgreen}{cmyk}{0.76,0,0.26,0.5}

\usepackage[a4paper]{geometry}
\usepackage{epsfig}

\usepackage{hyperref}
\hypersetup{
    bookmarks=true,         % show bookmarks bar?
    unicode=false,          % non-Latin characters in AcrobatÕs bookmarks
    pdftoolbar=true,        % show AcrobatÕs toolbar?
    pdfmenubar=true,        % show AcrobatÕs menu?
    pdffitwindow=false,     % window fit to page when opened
    pdfstartview={FitH},    % fits the width of the page to the window
    pdftitle={My title},    % title
    pdfauthor={Author},     % author
    pdfsubject={Subject},   % subject of the document
    pdfcreator={Creator},   % creator of the document
    pdfproducer={Producer}, % producer of the document
    pdfkeywords={keyword1} {key2} {key3}, % list of keywords
    pdfnewwindow=true,      % links in new window
    colorlinks=true,       % false: boxed links; true: colored links
    linkcolor=midnightblue,          % color of internal links (change box color with linkbordercolor)
    citecolor=magenta,        % color of links to bibliography
    filecolor=midnightblue,      % color of file links
    urlcolor=midnightblue,          % color of external links
}

\begin{document}

\title{ Emergent topological and half semimetallic Dirac Fermions at oxide interfaces} 

\author{Tian-Yi Cai}
\altaffiliation{Equal contribution.}
\affiliation{Department of Physics and Jiangsu Key Laboratory of Thin Films, Soochow University, Suzhou 215006, P. R. China }

\author{Xiao Li}
\altaffiliation{Equal contribution.}

\author{Fa Wang}
\affiliation{International Center for Quantum Materials, School of Physics, Peking University, Beijing 100871, P. R. China }

\author{Ju Sheng}
\email{ jusheng@suda.edu.cn}
\affiliation{Department of Physics and Jiangsu Key Laboratory of Thin Films, Soochow University, Suzhou 215006, P. R. China }

\author{Ji Feng}
\email{jfeng11@pku.edu.cn}
\affiliation{International Center for Quantum Materials, School of Physics, Peking University, Beijing 100871, P. R. China }

\author{Chang-De Gong}
\affiliation{Center for Statistical and Theoretical Condensed Matter Physics and Department of Physics, Zhejiang Normal University, Jinhua 321004, P. R. China }
\affiliation{National Laboratory of Solid State Microstructure and Department of Physics, Nanjing University, Nanjing 210093, P. R. China }

\begin{abstract}
It is highly desirable to combine recent advances in the topological quantum phases with technologically relevant materials. Chromium dioxide (CrO$_2$) is a half-metallic material, widely used in high-end data storage applications. Using first principles calculations, we show via interfacial orbital design that a novel class of half semi-metallic Dirac electronic phase emerges at the interface CrO$_2$ with TiO$_2$ in both thin film and superlattice configurations, with four spin-polarized Dirac points in momentum-space (k-space) band structure. When the spin and orbital degrees of freedom are allowed to couple, the CrO$_2$/TiO$_2$ superlattice becomes a Chern insulator  without external fields or additional doping. With topological gaps equivalent to 43 Kelvin and a Chern number 2, the ensuing quantization of Hall conductance to 2$e^2/h$ will enable potential development of these highly industrialized oxides for applications in topologically high fidelity data storage and energy-efficient electronic and spintronic devices.
\end{abstract}

\maketitle

The defining feature of massless Dirac Fermions \citep{Geim07, HasanKane10} is the linear energy dispersion, implying that a quasiparticle moves at a constant speed independent of its energy, as governed by relativistic quantum mechanics. The exploration of novel Dirac materials with \emph{bulk} Dirac or Dirac-Weyl states remains a great challenge in condensed matter and materials physics.\citep{Volkov85, Ishibashi08, Ran09, Wan11, Xu11, Asano11, Burkov11, Lu13} In particular, there has yet been neither theoretical proposal nor observation of materials systems with Dirac points a single spin species, which may be called a \emph{half-semimetal}. Half metals are a class of ferromagnetic materials in which the conduction is dominated by one spin component while the other spin is gapped, which stands to enable reduced device size and non-volatile memory. At a rudimentary level, a half semi-metal will harbour the fascinating Dirac physics \citep{Volkov85, HasanKane10, Geim07, Ishibashi08, Ran09, Wan11, Xu11, Asano11, Burkov11, Lu13} all with a single spin. And as a not-so-trivial consequence, it may lead to Chern insulators \citep{Haldane88, Qi06, Yu10, Chang13} or valleytronics materials \citep{Xiao07, Cao12, Xiao12, Li13} when the low-energy excitations turn massive. In particular, a Chern insulator exhibits quantum anomalous Hall (QAH) effect  in the absence of external magnetic field. This important fundamental phenomenon was established experimentally recently in magnetically doped topological insulator. \citep{Chang13} A spontaneous half semimetal will lead to a Chern insulator with appropriate spin-orbit coupling (SOC) and symmetry, which is advantageous for it requires no addition of impurity  to the system.

\begin{figure}[H]
\centering{}\includegraphics[width=13.5 cm]{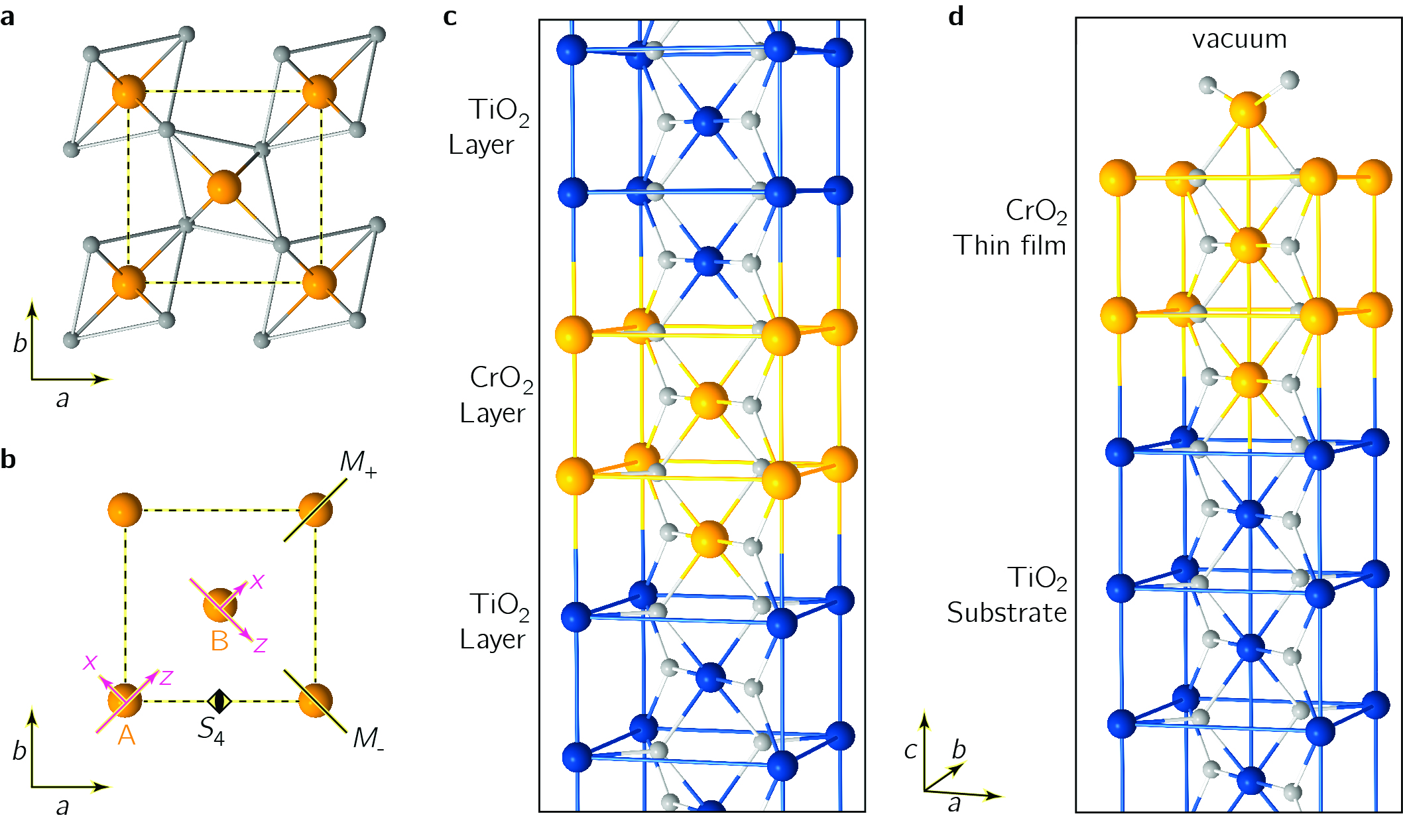}
\label{fig:structure}
\caption{Structure and symmetry of rutile oxides and the heterojunctions of CrO$_2$ and TiO$_2$. \textbf{a.} Rutile structure viewed down the $c$-axis. The XO$_6$ octahedra display two distinctive orientations, related to one another by the $4_2$ screw rotation. Octahedra of two orientations are indicated, the $C_2$ axes of which in this view are perpendicular to the plane. The X-O bonds along the crystallographic ${\bf a} \pm {\bf b}$ (or $[110]$ and $[1\bar{1}0]$) directions are elongated. Consequently, the local symmetry of the distorted octahedral shell is $D_{2h}$, for which the principal 2-fold axis may be chosen along the elongated X-O bonds. \textbf{b.} Symmetry and local coordinates of a XO$_2$ unit cell. Crystallographic $(1\bar{1}0)$ and $(110)$ planes are planes of mirror symmetry of the distorted octahedra, which we call $M_+$ and $M_-$, respectively. The system also has a four-fold rotoinversion, $S_4$. The \emph{local z} is aligned with $[110]$ for site A, whereas the \emph{local z} is aligned with $[1\bar{1}0]$ for site B. Hereafter, we will use $xyz$ for the local coordinates, and $abc$ for the global orientation. \textbf{c.} Structural model of superlattice of (CrO$_2$)$_4/$(TiO$_2)_{10}$ \textbf{d.} Structural model of CrO$_2$ thin film on TiO$_2$ substrate.  Orange, blue and gray balls correspond to Cr, Ti and O, respectively.}
\end{figure}

Precisely controlled growth of heterojunctions of oxides offers a unique avenue for exploring novel interfacial quantum phases, in which the interplay of orbital, charge, spin and lattice degrees of freedom gives rise to remarkably rich chemical, structural, electronic and magnetic varieties. Bulk CrO$_2$ (structure shown in Fig. 1\ref{fig:structure}) is a robust half metal even at elevated temperatures,\citep{Schwarz86, Kamper87, Dedkov02} and widely used in enterprise-grade data storage. Here, we consider the possibility to fashion the CrO$_2$ into single-spin Dirac states by interfacial orbital design. We combine theoretical and computational techniques to examine the interfacial half-metallic Dirac state at the CrO$_2$/TiO$_2$ junction. We show that by  forming a superlattice of CrO$_2$ and TiO$_2$, or  joining a thin film of CrO$_2$ to a TiO$_2$ substrate, the $d-$orbitals of Cr are isolated into interfacial Dirac states, with four single-spin Dirac points in the two-dimensional (2d) ${\bf k}$-space. Spin-orbit coupling (SOC) renders these Dirac nodes field-tunable; that is, they can continuously vary from massless to massive by aligning its magnetization. When the magnetization is perpendicular to the interfaces, the superlattice possess four massive Dirac nodes, each enclosing a $\pi$ Berry phase. The calculations indicate that this is a topological Chern insulator with Chern number = 2, which has twice the chiral edge states available from a Chern insulator based on topological insulators. \citep{Qi06, Yu10, Chang13}

\begin{figure}[H]
\centering{}\includegraphics[width=14.5cm]{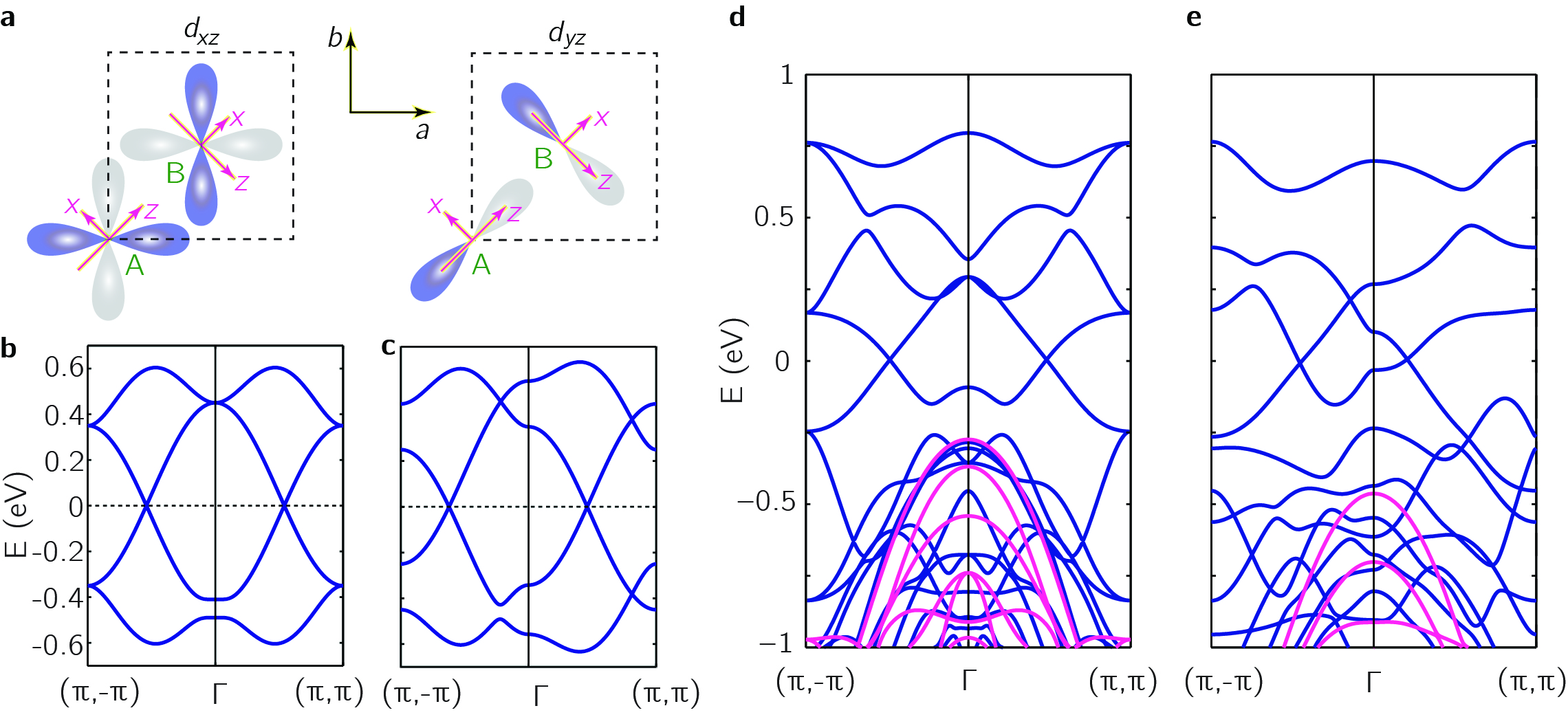}
\label{fig:model}
\caption{ Orbital design and single-spin Dirac points of CrO$_2$/TiO$_2$ interfacial states.
\textbf{a.} The Wannier-like basis orbital considered in our TB model. Left panel: $d_{xz}$ on sites A and B; Right panel: $d_{yz}$ on sites A and B. The red arrows represent the local coordinates. The lobes of orbitals are rendered with blue and gray shades, corresponding to a $\pi$ phase difference. \textbf{b.} TB band structure for the superlattice, with $S_4$ symmetry. \textbf{c.} TB band structure of the thin film model, absent $S_4$ symmetry. Parameters (all in eV): $t_1 = 0.3, t_2 = -0.3, t_3 = -0.01, t^{xz}_A = t^{xz}_B = -0.1, t^{yz}_A = t^{yz}_B = 0.1$;  $E^{xz}_{A} =  -0.05, E^{yz}_{A} = 0.05, E^{\alpha}_{B} = E^{\alpha}_{A}+\Delta_\mathrm{AB}$; For (b) $\Delta_\mathrm{AB} = 0$; and for (c) $\Delta_\mathrm{AB} =0.2 $. Bulk DFT calculations show that the $d_{xy}$ is energetically lower than $d_{xz},d_{yz}$. It follows that in the ferromagnetic ground state, the electrons will populate two of the four bands in our model. \textbf{d.} and \textbf{e.} show the non-relativistic DFT band structure of superlattice and thin film models, respectively. Here, the blue and red bands are two different spins. }
\end{figure}

Both CrO$_2$ and TiO$_2$ have stable rutile (XO$_2$, X = Ti, Cr) phases. In the rutile structure, each metal atom, Cr or Ti, lies in the center of an octahedral cage formed by six oxygen atoms. These XO$_6$ octahedra share corners to form a three-dimensional network, in the space group $P4_2/mnm$ (Fig. 1\ref{fig:structure}a). As shown in Fig. 1\ref{fig:structure}b, the set of symmetries, $M_\pm$ and $S_4$, constitutes the generators of the lattice point group. We now establish the bonding of a single CrO$_2$ bilayer cleaved along (001) plane, such as shown in Fig. 1\ref{fig:structure}a. The local symmetry at Cr sites is the $D_{2h}$ point group by axial elongation of the octahedron. The $t_{2g}$ orbitals are split into two sets, of which the ($d_{xz}, d_{yz}$) manifold is closest to the Fermi energy.\citep{Korotin98} Projection views of these two orbitals are shown schematically in Fig. 2\ref{fig:model}a. Each CrO$_2$ bilayer contains only two Cr in a tetragonal unit cell, referred to as sites A and B, respectively. A model can be constructed, taking into account four Wannier bases that transform as $d$-orbitals, $\{|d_{xz}(A)\rangle,|d_{yz}(A)\rangle,|d_{xz}(B)\rangle,|d_{yz}(B)\rangle \}$. A tight-binding (TB) spinless Hamiltonian, $\mathcal{H}_{\bf k}$, immediately follows, taking into account the nearest-neighbor (NN) and next nearest-neighbor (NNN) bonding,

\begin{subequations}
\begin{align}
\mathcal{H}_{\bf k} &= 
		\left[ \begin{array}{cc}
         H_A & H_{AB}\\
         H_{AB}^{\dagger} & H_B\end{array} \right] 
        ;\\
H_j &= 
\left[ \begin{array}{cc}
         \varepsilon^{xz}_j({\bf q}) & 0\\
        0 & \varepsilon^{yz}_j({\bf q}) \end{array} 
\right],\; j =A,B ;\\
H_{AB} &= 
\left[ \begin{array}{cc}
         -2 t_3 (\cos q_1 + \cos q_2)  & 2 i t_1 \sin q_1\\
        2 i t_2 \sin q_2  & 0 \end{array} 
\right].
\end{align}
\end{subequations}

\noindent where $q_{1,2} = (k_a \pm k_b)/2$, and $\varepsilon^\alpha_j({\bf q}) =  E_j^{\alpha} + 4 t^{\alpha}_j \cos q_1 \cos q_2$, with $-\pi\leq k_a,k_b<\pi$. Here, $E^\alpha_j$ are on-site energies and $t^\alpha_j$ the NNN hopping, with site index $j =$ A, B and orbital index $\alpha = d_{xz}, d_{yz}$. The NN hopping is described by $t_1, t_2, t_3$. Notice that $t_1$ and $t_2$ do not vanish because sites A and B are not on the same (001) plane. The NN hopping between $|d_{yz}\rangle$'s on A and B is zero, as ensured by $M_\pm$ mirror planes. 

The Hamiltonian proposed above leads to band structures shown in Figs. 2b and c. In Fig. 2b, the parameters are chosen respecting the $S_4$ symmetry; i.e., $E_A^\alpha = E_B^\alpha$ and $t^\alpha_A = t^\alpha_B$. We see that four Dirac points arise along the diagonals of the Brillouin zone (BZ) with equal distance away from the center of BZ. The four Dirac points are band contact points between the second and third Bloch bands. We may also break the $S_4$ symmetry, by setting unequal on-site energies of orbitals of A and B sublattices. In this case, four Dirac points are still present with unequal distances away from the center of BZ.  In both cases, all four Dirac points are located along the path of mirror planes, as dictated by symmetry, with $|k_{Da}|=|k_{Db}|$. Free-standing monolayer CrO$_2$ is quite unlikely to be fabricated experimentally. We therefore consider a thin film grown on a substrate. Thin film is necessary as the Dirac points will be obliterated in the case of the bulk CrO$_2$. And the substrate should favour epitaxial growth, such that the coherent interfacial bonding can be utilized to keep the opposite spin gapped away, preserving the single-spin Dirac points near the Fermi level. As shown in Figs. 1\ref{fig:structure}c and d, two types of (CrO$_2$)$_n$/(TiO$_2$)$_m$ heterostructures with indices (n-m), i.e. symmetric superlattices and asymmetric ultrathin films, are constructed based on the 2d CrO$_2$. Here, $m$ and $n$ refer to the number of XO$_2$ (X = Cr, Ti) formula units per unit cell in the model. To assess the stability, geometric and electronic structure of these heterostructures,  our $ab$ $initio$ calculations are performed using the projector augmented wave (PAW) method, as implemented in the Vienna \emph{ab initio} Simulation Package (VASP).\citep{Kresse96, Kresse99} They are based on density-functional theory (DFT) with the generalized gradient approximation (GGA) in the form proposed by Perdew, Burke, and Ernzerhof (PBE).\citep{Perdew98} On-site Coulomb interaction is included in the mean-field, rotationally invariant GGA+$U$ approach with $U-J=2$ eV for Cr $3d$ orbitals.\citep{Liechtenstein95,Korotin98} A plane-wave cutoff of 520 eV is used. The valence includes 12 electrons for Cr ($3p^{6}3d^{5}4s^{1}$), 12 for Ti ($3s^{2}3p^{6}3d^{2}4s^{2}$) and 6 for O ($2s^22p^4$). The in-plane lattice constant is set to 4.65 \AA, i.e. the optimized lattice constant of bulk TiO$_2$. For the ultrathin films, a vacuum layer larger than 16 {\AA} is adopted. We relax the ions towards equilibrium positions until the Hellman-Feynman forces are less than 0.01 eV/\AA.

As it turns out, the optimal $n = 4$ for the superlattice, and $n$ = 5 for thin film  heterostructure, as detailed in Figs. S1 and S2 in Supplementary Information (S.I. hereafter). These values give clear spin-polarized Dirac states at the Fermi level, absent other electronic levels in a 0.1 -- 0.2 eV window, as shown in Figs. 2\ref{fig:model}d and e. Similar to the case of bulk CrO$_2$, ferromagnetic (FM) ordering is energetically favoured in both systems. The $E_\mathrm{FM}-E_\mathrm{AFM} = -214$ and $-127$ meV/Cr for the superlattice and thin film, respectively. The non-magnetic configuration is more than 1.2 eV/Cr higher in energy. It is found that two bands cross the Fermi level at four distinct points along the zone diagonals, agreeing with the results from our TB Hamiltonian. The numerical gaps in the Kohn-Sham bands at the Dirac points are found to be less than 0.1 meV, indicating that the system is truly half semimetallic, and that the inter-CrO$_2$ layer coupling between periodic images through TiO$_2$ layers in the superlattice is vanishingly small. For the (4-10) superlattice, with the crossing points are located at ${\bf k}_{D} = \pm (0.482,\pm 0.482)\pi$. For the (5-9) thin film, they are located at $\pm (0.206, 0.206)\pi$ and  $\pm (0.454,-0.454)\pi$. Examining these two bands in the vicinity centered at the band crossing point ${\bf k}_{D}$, it is found that away from this point a gap opens linearly. These Dirac cones are rather anisotropic, with principal Fermi velocities falling in the range 1.0 -- 2.2 $\times 10^5$ m/s, roughly an order of magnitude smaller than that of graphene (see Fig. 3a). Analysis of the Kohn-Sham wavefunctions reveals that the dominant orbitals in these Dirac valleys are indeed combinations of $d_{xz}$ and $d_{yz}$ orbitals of different Cr atoms (see S.I.).

The Dirac points in the above interfacial electronic phase are quite extraordinary, from two points of view. First, they are single-spin Dirac species, unlike graphene that has two-fold spin degeneracy. Second, these Dirac points are field-tunable; that is, they can be massless (gapless) or massive (gapped) depending on straightforward field alignment. This tunability is associated with gap opening by SOC effect, as is introduced in non-self-consistent relativistic DFT calculations. When the crystallographic $c-$direction is the spin quantization axis, mirror symmetry is broken. Indeed, SOC lifts the degeneracy at the Dirac points, and a gap = 3.7 meV is opened up in the (4-10) superlattice (see Figs. 3\ref{fig:chern}a -- b). To ascertain that the gap is induced by SOC, and is accurately determined, we examine how the gap depends on the strength of SOC. As shown in Fig. \ref{fig:chern}3c, the linearity of gap vs. SOC strength plot gives accurate extrapolation to the band gap at full speed of light, assuring that the gap is induced by SOC and is computed correctly. In the case of (5-9) thin film, SOC opens gaps differently for Dirac points on the diagonal ($k_a= k_b$) and antidiagonal ($k_a = -k_b$) of the BZ. The gap is  over 5 meV on the antidiagonal, but about 0.5 meV on the diagonal. The thin film remains semimetallic by the electron-hole compensation between valleys (see S.I.). On the other hand, if the superlattice is magnetized by an in-plane field, one expects a momentum shift of the Dirac point rather than gap opening.\citep{Zhang13} Indeed, we have computed the electronic structure of CrO$_2$/TiO$_2$ superlattice with spin quantization axis along crystallographic [110] or $[1\bar{1}0]$. As these Dirac points are located at general points, it is numerically difficult to locate the exact degeneracy. We found with very dense ${\bf k}$-mesh that the gaps in this case are less than 0.1 meV in all valleys. Fig. 3d  further shows the band gap of the superlattice gradually increases as the magnetization direction rotates from [110] to $c-$direction, demonstrating that the Dirac states of the interfacial quantum phase can be continuously tuned by rotating its magnetization.

\begin{figure}[H]
\centering{}\includegraphics[width=8.5 cm]{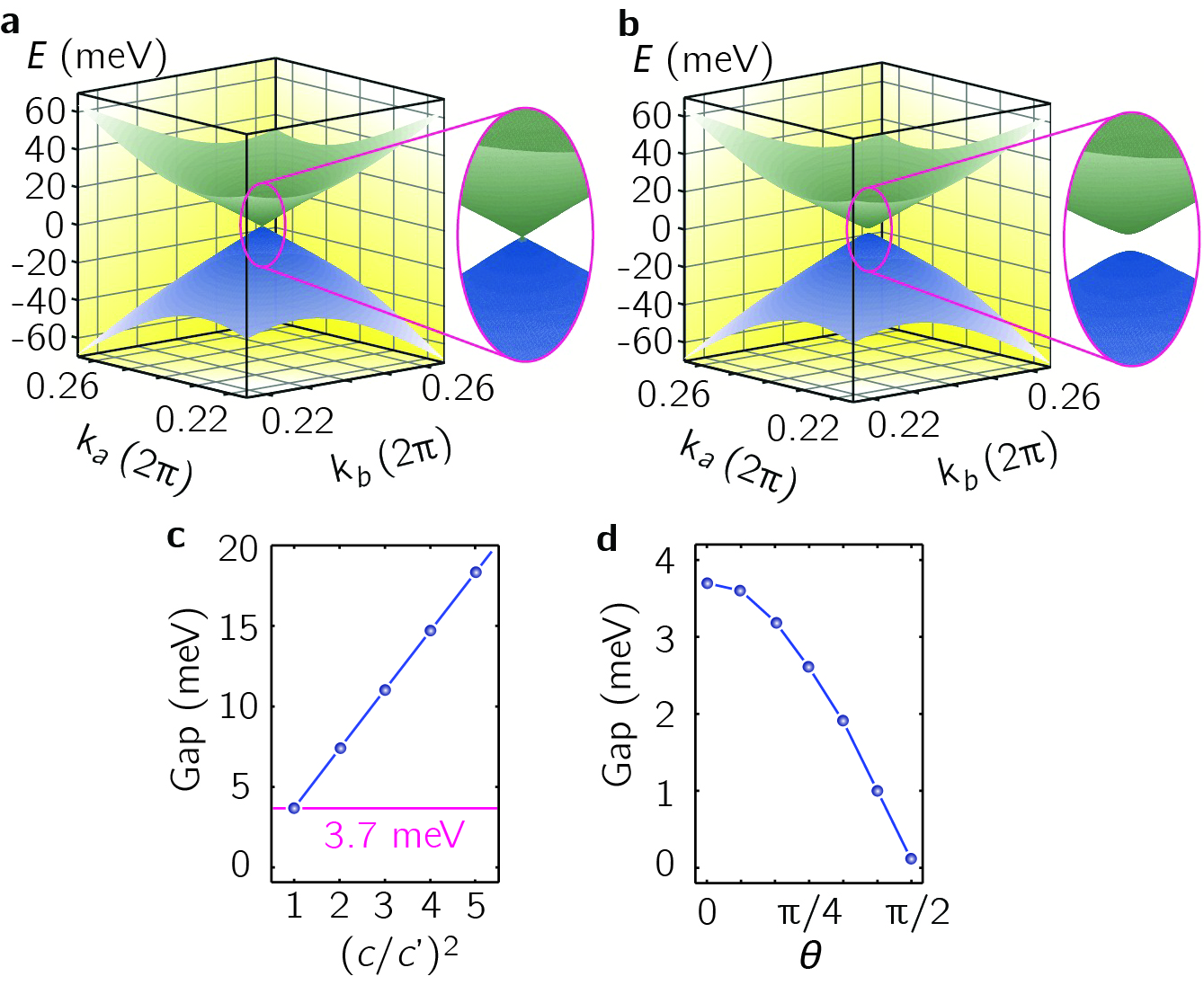}

\caption{ Field-tunable Dirac points in (CrO$_2)_4$/(TiO$_2)_{10}$ superlattice.
\textbf{a.} The 2d band structure without SOC in the vicinity of a Dirac point in the superlattice. \textbf{b.} The 2d band structure with SOC in the vicinity of a Dirac point in the superlattice, using the crystallographic $c-$direction as the spin quantization axis. A finite band gap is clearly opened. \textbf{c.}  The band gap of the superlattice as a function of the strength of SOC, with the spin quantization axis along $c-$direction. The strength of SOC is characterized by $(c/c')^2$, where $c$ is the speed of light \emph{in vacuo}, and $c'$ the artificially retarded speed of light used in actual calculations. \textbf{d.} The band gap of the superlattice with rotating the direction of magnetization from crystallographic [110] to [001] direction within the $(1\bar{1}0)$ plane. The angle made by [001] direction and the spin quantization axis is denoted $\theta$.  }
\end{figure}

The gaps opened in the superlattice by [001] magnetization are related by symmetry, which actually ensures its unusual topological character. These single-spin Dirac points are located along low-symmetry lines in the BZ, in contrast with typical Dirac systems such as graphene and the surface states of 3-dimensional topological insulators, of which the Dirac points are found at high-symmetry points. These observations bear non-trivial consequences. If the system breaks the mirror symmetry, both $M_+$ and $M_-$, in the magnetic group, but preserves the $S_4$ symmetry, it is guaranteed to be a non-trivial Chern insulator. It is, therefore, of interest to quantify the topological character of the fully SOC-gapped superlattice, by computing the $U(N)$ Berry curvature of Kohn-Sham Bloch states. The Berry curvature, $\Omega({\bf k})$, is computed directly from the Kohn-Sham Bloch wavefunctions on a discretized ${\bf k}$ mesh, as\citep{KingSmith93,Hatsugai05}
\begin{equation}
\Omega({\bf k}) \delta k_a \delta k_b = \mathrm{Arg \{ det[} 
L_a({\bf k}) L_b({\bf k}+\delta \hat{\bf k}_a) 
L_a^\dagger({\bf k}+\delta \hat{\bf k}_b)
L_b^\dagger({\bf k})]\}
\end{equation}
where $L_\alpha({\bf k}) \equiv \vec{u}^\dagger({\bf k}) \cdot \vec{u}({\bf k} + \delta \hat{\bf k}_\alpha)$ with $\alpha = a,b$ and $\delta \hat{\bf k}_\alpha = \delta k_\alpha\hat{\bf e}_\alpha$. Here, $\vec{u}({\bf k}) = [|u_1({\bf k})\rangle, ..., |u_N({\bf k})\rangle]$, with $u_j({\bf k})$ being the cell-periodic part of the Bloch wave function, $\psi_j({\bf k})$, of the $j$th band at ${\bf k}$. The ${\bf k}$-periodic gauge is imposed such that $\psi_j({\bf k}+{\bf K})=\psi_j({\bf k})$, where ${\bf K}$ is a reciprocal lattice vector. A quarter BZ, $k_a,k_b \in [0,\pi]$, is pixelated into $60\times60$ plaquettes for evaluation of $U(N)$ Berry connections. A total of 176 bands up to the band gap are considered, below which there is a sub-valence gap of about 10 eV. The full zone Berry curvature (see Fig. 4a) can be retrieved by imposing $S_4$ symmetry. Numerically integrating the non-Abelian Berry curvature over the 2-torus ($\mathbb{T}^2$) BZ yields a Berry phase of $4.00 \pi$. We note that the Hall conductance, $\sigma_{xy} = (1/2\pi S)\int d{\bf k} \Omega$, where $S$ is the area of a unit cell and $\sigma_{xy}$ is given in the units of conductance quantum, $e^2/h$.\citep{Thouless82} It is concluded that the interfacial material is a Chern insulator and will exhibit a fully quantized Hall conductance, $\sigma_{xy} = \pm 2.00 \;e^2/h$. Here, the $\pm$ sign corresponds to the sense of magnetization of the material prior to the Hall measurements.  

Experimentally, this system offers a few advantages over the only experimentally established QAH system, which required magnetic doping of topological insulators and showed full quantization of Hall conductance at about $0.03$  K.\citep{Chang13} First, the CrO$_2$/TiO$_2$ is ferromagnetic spontaneously, and therefore does not require additional magnetic impurity atoms, which are a source of scattering and could have lowered the transition temperature. Second, the band gap of the present system corresponds to 43 K, indicating that full quantization of the Hall conductance may be achieved at considerably higher temperatures than the magnetically doped bismuth chalcogenides.\citep{Chang13} Third, the present system has unique symmetry that ensures its topological character and is composed of two highly industrialized oxides, for which process technologies are rather mature for fabrication of the heterostructure.\citep{Kamper87, Dedkov02} 

\begin{figure}[H]
\centering{}\includegraphics[width=15.2 cm]{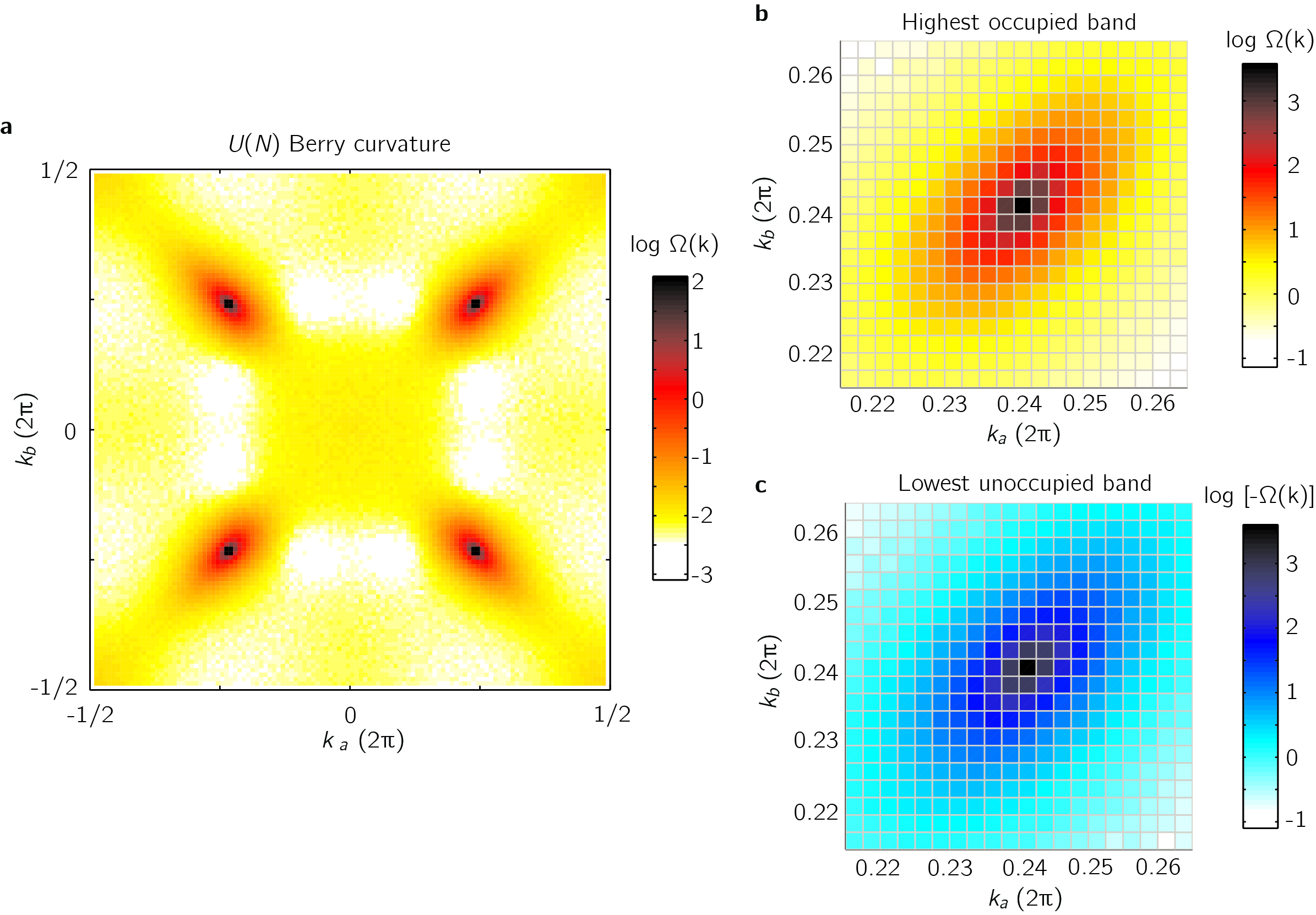}
\label{fig:chern}
\caption{ Topological character of the (CrO$_2)_4$/(TiO$_2)_{10}$ superlattice interfacial states.
\textbf{a.} The non-Abelian Berry curvature over the entire BZ. Notice that the Berry curvature is given in the units of area of real-space unit cell, 21.623 \AA$^2$, and is viewed on a base-10 logarithmic scale.
\textbf{b.} and \textbf{c.} The $U(1)$ Berry curvature of the massive Dirac valley (\emph{c.f.} Fig. 3\textbf{b}) of the highest occupied valence band and the lowest unoccupied conduction band. Notice that the $-\Omega$ is plotted for the conduction band. Each plot contains $20 \times 20$ square tiles, each corresponding to a plaquette around which Berry connections are computed.   }
\end{figure}

Fig. 4a also shows that the main contribution to the $U(N)$ Berry curvature in the (4-10) superlattice is sharply concentrated around the locations of the four massive Dirac states, indicating that the main source of the anomalous conductivity arises from the SOC gaps. To confirm this, we compute the the $U(1)$ Berry curvatures of the valence and conduction bands across the gap, as shown in Figs. 4b and 4c.  We see that the massive Dirac fermion states show prominent $U(1)$ Berry curvature near the gap, with opposite signs for the electron and hole excitations. The integration of the Berry curvature of the highest occupied band in the vicinity of the gap over the small ${\bf k}-$space region for a single valley encloses a Berry phase $\sim 0.95 \times \pi$. By the $S_4$ symmetry of the system, the four valleys will have identical contribution to the Hall conductance. This confirms that the SOC gap dominates the anomalous velocity of Bloch electrons. 

In summary, we demonstrated that the interfacial electrons at the CrO$_2$/TiO$_2$ heterojunction can be fashioned into single-spin Dirac states. In the superlattice model, we can obtain a Chern number = 2 QAH phase, which will enable dissipationless digital signal transmisstion with minimal influence of noise. The first-order experiments will include spin-polarized ARPES to characterize these field-tunable Dirac states. CrO$_2$/TiO$_2$ heterostructures will  behave more like a single-spin massless Dirac semimetal with in-plane magnetization, which, for example, will exhibit unconventional quantum Hall effect \citep{Geim07} with single spin. Direct Hall measurements on the superlattice samples with out-of-plane magnetization will reveal the conductance plateau \citep{Yu10, Chang13}, which should be robust against weak disorder that may be present in the sample. For both systems, different substrates, instead of TiO$_2$, with heavier elements should also be assayed, as SOC of the substrate may further boost the band gap via proximity effect (\emph{c.f.} Fig. 3c). 

\bigskip

\noindent {\em Acknowledgements}. The authors are grateful for useful discussions with Profs. Guang-Yu Guo and Junren Shi. This work is supported by the National Science Foundation of China under Grants Nos. 11374220, 11174009 and 11104193, by China 973 Project No. 2013CB921900, and by Priority Academic Program Development of Jiangsu Higher Education Institutions.

\end{document}